\documentclass[aps, pra, twocolumn,showpacs,floatfix]{revtex4-1}
\usepackage[T1]{fontenc}
\usepackage[latin9]{inputenc}
\usepackage{amssymb}
\usepackage{times}
\usepackage{graphicx}
\usepackage{amsmath}
\usepackage{pstricks}
\usepackage{color}
\usepackage{subfigure}
\usepackage{bm}

\newcommand{\qu}[1]{``#1''}
\newcommand{\bb}[1]{\left(#1\right)}                                		% (..)
\newcommand{\meanv}[1]{\left\langle#1\right\rangle}							% <...>
\newcommand{\intrp}{\int  \mathrm{d}^3 \,r^{\prime}} 	% integral
\newcommand{\intr}{\int  \mathrm{d}^3 \,r} 	% integral
\newcommand{\intI}[2]{\int #2\! \mathrm{d} \,#1} 	% integral
\newcommand{\psiC}{\Psi\bb{\bm{r}}}                        % hat Psi(r)
\newcommand{\psiCp}{\Psi\bb{\bm{r}^{\prime}}}                        % hat Psi(r)
\newcommand{\vdip}{V_{dip} \bb{\bm{r} }}
\newcommand{\vdipDif}{V_{dip} \bb{\bm{r} - \bm{r}^{\prime}}}
\newcommand{\ii}{\dot{\imath}}
\newcommand{\ki}[1]{\bm{k}_{\bm{#1}}}

\begin{document}
\title{Correlations of quasi-2D dipolar ultracold gas at finite temperatures}

\author{Krzysztof Paw{\l}owski$^{1,2}$, Przemys\l aw Bienias$^{1, 2}$, Tilman Pfau$^2$, Kazimierz Rz\k a\.zewski,$^{1,2}$}
\affiliation{\mbox{$^1$Center for Theoretical Physics PAN, Al. Lotnik\'ow 32/46, 02-668 Warsaw, Poland}  \\
\mbox{$^2$5. Physikalisches Institut, Universit\"at Stuttgart, Pfaffenwaldring 57, 70550 Stuttgart, Germany}}

\begin{abstract}
We study a quasi two dimensional dipolar gas at finite, but ultralow temperatures using the classical field approximation.
The method, already used for a contact interacting gas, is extended here to samples with a weakly interacting long-range inter-atomic potential.
%We compute the stability diagram at increasing temperatures and contrast it with the zero temperature case analyzed with the help of Bogoliubov analysis.
%Furthermore 
We present statistical properties of the system for the current experiment with Chromium \cite{mueller2011} and compare them with statistics for atoms with larger
magnetic dipole moments. Significant enhancement of the third order correlation function, relevant for the particle losses, is found.
\end{abstract}

%\pacs{67.85.Hj, 05.30.-d}

% 03.50.Kk  Other special classical field theories 
% 05.30.-d Quantum statistical mechanics (for quantum fluids aspects, see 67.10.Fj)
% 03.75.-b Matter waves 
% 67.85.-d Ultracold gases, trapped gases
% 67.85.Hj Bose-Einstein condensates in optical potentials

\maketitle
%%%%%%%% INTRO
The successful cooling down of dipolar gases below the condensation temperature \cite{griesmaier2005} 
has attracted many theorists and experimental groups.
Dipolar forces may stabilize or destabilize the ultracold cloud depending on its polarization. A collapsing cloud forms
 so called Bose-Nova \cite{lahaye2008}. Dipolar forces also introduce inter-site effects between atoms loaded in optical lattices \cite{mueller2011}.
From the theory side, a variety of new phenomena are expected, i.e. new quantum phases \cite{goral2002} or rotonization of the spectrum  \cite{santos2003}.
The interest of the community is still increasing, as the new atomic species, Erbium \cite{aikawa2012} and Dysprosium \cite{Lu2011},
has been condensed recently. Furthermore, a big effort has been made to condense polar molecules with electric dipole interactions
orders of magnitude larger than the magnetic ones \cite{aikawa2010, Ni2008}.
%heternouclear molecules  \cite{Ni2008}

Especially interesting is a quasi two dimensional system, where roton may be formed \cite{ronen2007}. 
The latter, although investigated in many theoretical papers, has not been observed in dipolar gases yet.
On the other hand it is known, that the two dimensional system at finite temperatures has many interesting features even in
a purely contact interacting gas. 
The most intriguing is the Berezinsky-Kosterlitz-Thouless transition, also studied experimentally \cite{hadzibabic2006}.

Thus, it is natural to look at a two dimensional dipolar gas at temperature above the absolute zero.
The method applied most often for such study is the Bogoliubov approximation  \cite{ronen2007}.
Such treatment neglects higher order interactions between modes and is limited to low temperatures only.
Moreover in the case of an untrapped system, it assumes a condensation in the zero-momentum mode, whereas close to instability this
is not necessarily the case. 
For system in a trap the appropriate shape of the condensate and, partially, interactions between excitations have been included in extensions of the Bogoliubov approximation
 \cite{ticknor2012, ticknor2012B, bisset2012}, however, at the cost of more computationally expensive algorithms. 
Another qualitative approach is based on the mean field method \cite{bisset2011}.

Among  other very promising techniques  are the classical field methods (called also $\textit{c}$-field methods).
In the case of a trapped system the numerical algorithm for a projected Gross-Pitaevskii equation has already been  presented \cite{blakie2009}.

We present an even simpler example of the $\textit{c}$-field methods' family. 
It has been successfully used for deriving the full statistics in a thermal equilibrium of an ideal gas \cite{witkowska2009, witkowska2010} 
and a contact interacting gas for weakly repulsive \cite{bienias2011, karpiuk2012}, as well as attractive forces \cite{bienias2011a}. 
Here, we apply this method to the dipolar gases, trapped in a "hybrid" geometry - a strong harmonic trap in the $z$ direction 
and a box with periodic boundary condition in the perpendicular directions. 
With our method we gain qualitative insight into the statistical properties of the system, without any ad-hoc assumption for a 
shape of the mostly occupied state and 
including interactions in all orders.
We use the canonical ensemble, thus the total number of atoms is conserved.

The paper is organized as follow: 
In Sec. \ref{sec:methods} we sketch our realization of the classical field approximation (CFA). 
The derivation of the dipolar interaction energy functionals, has been relegated to the Appendix. 
%In Sec. \ref{sec:diagram} we discuss the stability diagram contrasted with solutions from Bogoliubov approximation. 
Sec. \ref{sec:experimental_conditions}  focuses on the analysis of the, up to date, the best experimental
realization of quasi 2D dipolar gases, done with Chromium in \cite{mueller2011}, and compare it with Erbium and Dysprosium. 
The most important results are summarized in Conclusions \ref{sec:conclusion}.

\section{Classical field approximation\label{sec:methods}}
We study a weakly interacting dipolar Bose gas trapped in a very steep harmonic potential in the $Z$ direction and square box with periodic boundary condition in the $X$ and $Y$ directions.
Thus, our Hamiltonian of the three dimensional Bose gas has a form:
\begin{eqnarray}
\nonumber H&=& \intr \,\hat{\Psi}^{\dagger }(\bm{r})\left( \frac{\hat{p}^2}{2m}+\frac{1}{2}m\,\omega_z^2 z^2  \right) \hat{\Psi}(\bm{r}) +\\
 & +&\frac{g}{2} \intr \,\hat{\Psi}^{\dagger }(\bm{r}) \hat{\Psi}^{\dagger }(\bm{r}) \hat{\Psi}(\bm{r}) \hat{\Psi}(\bm{r}) +\\
\nonumber  &+&\frac{1}{2} \intr\,\mbox{d}^3r^{\prime} \,\hat{\Psi}^{\dagger }(\bm{r}) \hat{\Psi}^{\dagger }(\bm{r^{\prime}}) \vdipDif \hat{\Psi}(\bm{r^{\prime}}) \hat{\Psi}(\bm{r}).
\label{eqn:hamiltonian}
\end{eqnarray}
The Hamiltonian is a sum of a single particle energy, a conventional contact interaction energy with the coupling constant $g$ and a dipolar interaction energy.
The dipolar potential is given by:
 \begin{equation}
 \vdip = \frac{3g_{dd}}{4\pi}\frac{r^2-3\bb{\bm{n} \bm{r}}^2}{r^5},
\label{eqn:vdip}
\end{equation}
where $\bm{n}$ is a unit vector in the direction of polarization and $g_{dd}$ parameterizes 
%a measure of 
  strength of dipolar interactions.
The periodic boundary conditions in the $X$, $Y$ directions are responsible for the quantization of momenta:
\begin{equation}
\bm{k}_{\bm{i}}  = \frac{2\pi}{L} [i_x, i_y],\quad \bm{k} = [\bm{k}_{\bm{i}}, k_z]
\label{eqn:discrete-momenta}
\end{equation}
where $L$ is the size of the box and $i_x, i_y $ are any integer numbers. Here we use the symbol $\bm{i}$ to denote the two dimensional index.
The two dimensional vector $\bm{k}_{\bm{i}}$ is a projection of a total momentum $\bm{k}$ onto $XY$-plane, whereas $k_z$ is the $Z$ component of $\bm{k}$.
We restrict our considerations to such strong frequencies $\omega_z$, small interaction energies and low temperatures, 
that the shape of the atomic cloud does not change in the $Z$ direction. 
In other words we assume, that a convenient base is provided by the product of the ground state Gaussian function
 in the strongly confined $Z$ direction
\begin{equation}
 \psi_0 (z) = \bb{ \frac{1}{\pi l_z^2}} ^{1/4} \exp \bb{-\frac{z^2}{2l_z^2}}
\end{equation}
with $l_z$ being the Gaussian width (dispersion), 
and plane waves in other directions.
Thus the suitable expansion of the field operator is given by:
\begin{equation}
\hat{\Psi}(\bm{r}) = \frac{1}{L}\sum _{\bm{i}} \psi_0 (z) e^{-\ii \bm{\rho} \bm{k}_{\bm{i}}}\hat{a}_{\bm{i}} ,
\label{eqn:fieldOp}
\end{equation}
where $\hat{a}_{\bm{i}}$ annihilates an atom in the $\bm{i}$-th momentum state, and $\bm{r} = (\bm{\rho}, z)$.

The classical field approximation consists of replacing the creation and annihilation operators by classical complex amplitudes:
\begin{equation}
\hat{a}_{\bm{i}}, \hat{a}_{\bm{i}}^{\dagger} \mapsto \sqrt{N}\alpha_{\bm{i}}, \sqrt{N}\alpha_{\bm{i}} ^{*}
\label{eqn:mapowanie}
\end{equation}
It has been shown in \cite{witkowska2009}, that probabilistic properties of the condensate for two dimensional ideal Bose gas in a box 
are for large $N$ perfectly reproduced by the classical fields approximation 
provided the number of degrees of freedom is kept finite with the last retained state chosen as:
\begin{equation}
\hbar^2\bm{k}_{max}^2/2m = 0.68k_B T
\label{eqn:kvst_g0}
\end{equation}
where $T$ is the absolute temperature and $k_B$ is the Boltzmann constant. 
However  for small systems, the formula \eqref{eqn:kvst_g0} needs a significant correction.
In general, the optimal cut-off momentum $\bm{k}_{max}$ 
is obtained from fitting the distribution of the number of condensed
atoms from the classical field to the exact solution in the ideal gas case. 
The latter is obtained according to  \cite{wilkens1997}, 
the former can be also exactly calculated as described in the Appendix of  \cite{bienias2011}.
For instance, in the system described in the next section, the appropriate cut-off is given by:
\begin{equation}
\hbar^2\bm{k}_{max}^2/2m = 0.99k_B T,
\label{eqn:kvst}
\end{equation}
a result which is obtained by a direct calculation.
This version of classical fields approach was successfully used for the weakly contact interacting gas \cite{witkowska2010, bienias2011, bienias2011a}.
It also provided a quantitative description of recent measurements (supplemental material in \cite{karpiuk2012}).

The energy of the complex amplitudes configuration for dipolar gas is given by:
\begin{equation}
E\left( \left\{ \alpha_i \right\} \right) = \sum_{\bm{i}}^{K_{max}} \frac{\hbar^2 \ki{i}^2}{2m} |\alpha_{\bm{i}} |^2 + E_{int}\left( \left\{ \alpha_i \right\} \right),
\label{eqn:energy0}
\end{equation}
where $N$ is the total number of atoms and $E_{int}\left( \left\{ \alpha_i \right\} \right)$ is the quartic polynomial in the amplitudes $\left\{\alpha_i\right\}$.
The sum is over all indices $\bm{i}$, which obey inequality $i_x^2 + i_y^2<K_{max}^2$. 
The cut-off $K_{max}$ is the index of the last retained mode, defined as $2\pi K_{max}/L = k_{max}$.
The interaction energies (dipolar and contact together) are given by
\begin{equation}
 E_{int} = E_{dip}^{II} + \frac{\bar{g} Nn_{2D} }{2\sqrt{2\pi}l_z}\sum_{\bm{m}, \bm{p}, \bm{n}}^{K_{max}} \alpha_{\bm{m}}^*\alpha_{\bm{p}}^*\alpha_{\bm{n}}\alpha_{\bm{m}+\bm{p}-\bm{n}},
\label{eqn:intE}
\end{equation}
where $n_{2D}=N/L^2$ is the two dimensional density.
Here we introduced an effective contact interaction coupling constant $\bar{g} =g + g_{dd}\bb{\cos 2\theta + \cos^2\theta}$, where
$\theta$ is an angle between the direction of the dipoles' polarization and the $Z$ axis. 
The second part of the dipolar interaction $E_{dip}^{II}$ consists of the long range potential between dipoles:
\begin{equation}
E_{dip}^{II} =   \frac{g_{dd}n_{2D}N}{l_z} \sum_{\bm{m}, \bm{p}, \bm{n}}^{K_{max}} \alpha_{\bm{m}}^*\alpha_{\bm{p}}^*\alpha_{\bm{n}}\alpha_{\bm{m}+\bm{p}-\bm{n}} \tilde{V}^{2D}(\ki{n}-\ki{m}),
\label{eqn:edipLongRange}
%E_{dip}^{II} =   \frac{g_{dd}n_{2D}N}{l_z} \sum_{\bm{m}, \bm{p}, \bm{n}}^{K_{max}} \alpha_{\bm{m}}^*\alpha_{\bm{p}}^*\alpha_{\bm{n}}\alpha_{\bm{m}+\bm{p}-\bm{n}}T_{mpn} \tilde{V}^{2D}(\ki{n}-\ki{m}),
\end{equation}
with
\begin{equation}
\tilde{V}^{2D}(\bm{k}) =-\frac{3l_z}{4} \frac{k_x^2\cos 2\theta + k_y^2\cos^2\theta}{k}  e^{k^2 l_z^2/2} \mathrm{Erfc} \bb{\frac{kl_z}{\sqrt{2}}} .
\label{eqn:vdip2Dv2}
\end{equation}
% Here for shorter notation we defined 
% \begin{equation}
% T_{mpn} = \sum_{\bm{m}, \bm{p}, \bm{n}}^{K_{max}} \alpha_{\bm{m}}^*\alpha_{\bm{p}}^*\alpha_{\bm{n}}\alpha_{\bm{m}+\bm{p}-\bm{n}},
% \label{eqn:tmnp}
% \end{equation}
%  which depends only on the set $\{\alpha_i\}$.
The derivation of formulas \eqref{eqn:intE}-\eqref{eqn:vdip2Dv2}, with a short discussion of the potential \eqref{eqn:vdip2Dv2}
is presented in details in the Appendix \ref{appendix:DipolarInteraction}. 

The initial three dimensional problem is reduced to a two dimensional one. Hence the classical field 
\begin{equation}
\Psi (\bm{\rho} ) = \int \mbox{d}z\,\Psi (\bm{r} )  \psi_0(z)
\end{equation}
will refer from now on
%this point 
to a two dimensional function.

From Eq. \eqref{eqn:intE} it is clear that the dipole interaction in the geometry considered here partially acts like a contact interaction, thus it shifts the coupling constant $g$. 
Furthermore, this shift in the coupling constant may be tuned by the polarization direction. 
At the angle  $\theta\approx 0.96$, at which $\cos 2\theta + \cos^2\theta = 0$, this shift vanishes.
\begin{figure}
 \includegraphics[width=6cm]{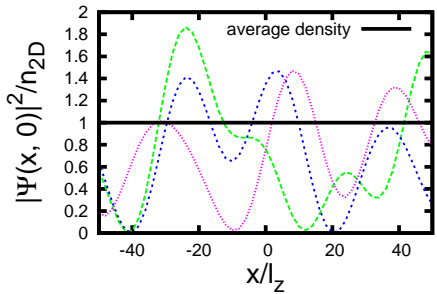}
\caption{(Color online) Cuts through single exemplary copies of atomic density (three non-solid lines) 
obtained during generation of the canonical ensemble with the Metropolis algorithm, 
together with the average density over the ensemble (solid black line).
\label{fig:densityMetropolis}}
\end{figure}

At finite temperatures the probability distribution of a given configuration of the classical field according to the canonical ensemble is:
\begin{equation}
P\left( \left\{ \alpha_i \right\} \right) = \frac{1}{Z(N,T)} \exp \left[-\frac{E\left( \left\{ \alpha_i \right\} \right)}{k_B T}\right] ,
\label{eqn:probability}
\end{equation}
where $Z(N,T)$ is the classical partition function for $N$ atoms at temperature $T$. 

The average of any observable $\hat{\mathcal{O}}$ is computed according to:
\begin{equation}
 \meanv{\mathcal{O}} = \int \mbox{d} \alpha_0\ldots\mbox{d} \alpha_{\bm{k}_{max}} P\left( \left\{ \alpha_i \right\} \right) \mathcal{O}\left( \left\{ \alpha_i \right\} \right),
\end{equation}
where integrals go through all the possible complex amplitudes subjected to the constraint forced by the fixed number of particles:
\begin{equation}
\sum^{K_{max}}_{{\bm{i}}}|\alpha_{\bm{i}}|^2=1.
\label{eqn:normalization}
\end{equation}
The best known Monte Carlo realization of this distribution is given by the Metropolis algorithm \cite{metropolis1953}. 
With it we explore the multidimensional phase space built from a set of $\left\{\alpha_i\right\}$ as described in detail in \cite{witkowska2009}.

The algorithm realizes the idea of a statistical ensemble - it looks for suitable copies of the system (different configurations $\left\{ \alpha_i \right\}$), 
all copies with energies distributed according to the Boltzmann distribution. 
As illustrated in Fig. \ref{fig:densityMetropolis} the single elements of the ensemble often differ significantly from the average of them.
Typically, computations require millions of generated copies to reach reasonably smooth averages. 

Following \cite{penrose1956}, the identification of the condensate requires a diagonalization of the single particle density matrix:
\begin{equation}
\rho_{i,j}=\langle\alpha_i^* \alpha_j\rangle=\sum_n\lambda_n\, \beta_i^*(n) \beta_j(n)
\label{eqn:rho}
\end{equation}
where the mean value is taken with the probability distribution \eqref{eqn:probability} 
and the eigenvector $\beta (0)$ corresponding to the dominant eigenvalue $\lambda_0$ (if such one exists) is the condensate.

Let us show the outcome of the method for a special
case of very low temperature.
As the global phase has no physical meaning, we can always eliminate $\alpha_{\bm{0}}$ choosing it as real and
 equal to $\alpha_{\bm{0}} = \sqrt{1 - \sum_{\bm{k} \neq 0} |\alpha_{\bm{k}}^* \alpha_{\bm{k}}|}$.
The interaction energy in \eqref{eqn:energy0} will remain a quartic
polynomial in the amplitudes $\left\{ \alpha_{\bm{q}\neq 0} \right\}$. 
If the amplitude of the zero momentum
component is dominant, that is $|\alpha_0| \approx 1\gg |\alpha_{\bm{k}\neq 0}|$
we can follow  the derivation of the Bogoliubov approximation. 
It requires neglect of the third and the fourth order
terms in the interaction energy. The remaining quadratic form is
easily diagonalized yielding:
\begin{equation}
E\bb{ \left\{\alpha_{\bm{k}}\right\}} = E_0 + \sum_{\bm{k} \neq 0} E_{\bm{k}} \beta^{*}_{\bm{k}}\beta_{\bm{k}},
\end{equation}
where $E_0$ is a constant, the spectrum $E_{\bm{k}}$ reads
\begin{equation}
E_{\bm{k}} =  \sqrt{\frac{\hbar^2q^2}{2m} \bb{\frac{\hbar^2q^2}{2m} + 
\frac{4n_{2D}}{l_z}\bb{
\frac{ \bar{g} }{ 2\sqrt{2\pi} } + g_{dd}   \tilde{V}^{2D} (\bm{k})
} } }
\label{eqn:bogol}
\end{equation}
and the new complex amplitudes $\beta_{\bm{k}}$ are defined as
\begin{eqnarray}
\beta_{\bm{k}} &= & u_{\bm{k}} \alpha_{\bm{k}} - v_{\bm{k}} \alpha_{-\bm{k}}^* \\
\nonumber u_{\bm{k}} + v_{\bm{k}} & = & \bb{\frac{\frac{\hbar^2k^2}{2m}}{\frac{\hbar^2k^2}{2m} + \frac{2n_{2D}}{l_z}\bb{
\frac{ \bar{g} }{ 2\sqrt{2\pi} } + g_{dd}   \tilde{V}^{2D} (\bm{k})
}}}^{1/4}\\
\nonumber u_{\bm{k}} - v_{\bm{k}} & = &\bb{\frac{\frac{\hbar^2k^2}{2m}}{\frac{\hbar^2k^2}{2m} + \frac{2n_{2D}}{l_z}\bb{
\frac{ \bar{g} }{ 2\sqrt{2\pi} } + g_{dd}   \tilde{V}^{2D} (\bm{k})
}}}^{-1/4} .
\end{eqnarray}

Thus the Bogoliubov spectrum, presented for instance in  \cite{blakie2012, sykes2012}, appears here as a special case.
\section{Experimental conditions\label{sec:experimental_conditions}}
Here we present the statistics of the quasi 2D dipolar gas for parameters close to the experimental ones, used in \cite{mueller2011}.
The experiment has been done in a deep optical lattice at temperatures around $100$nK. In each lattice site, the cloud was indeed quasi2D 
(checked with the help of numerically solved three dimensional Gross-Pitaevskii equation) with $\omega_z /2 \pi = 36573$Hz.

Our calculations are made for such parameters, that mimic the conditions in a single central lattice site.
Thus we keep the aspect ratio between the Gaussian width in the $Z$ direction and the size of the system in others equal to $100$.
To ensure a density of atoms close to the experimental value ($\approx 10^{14}$/cm$^3$) we consider here $N=500$ atoms.
We characterize the strength of the dipolar interaction with the "dipolar length" $a_{dd} = \frac{mg_{dd}}{4\pi\hbar^2}$  \cite{koch2008}, 
which for Chromium equals  $16a_B$, where $a_B$ is the Bohr radius.
All results are presented in oscillatory units, i.e. length is given in $l_z=\sqrt{\hbar/m_{Cr}\omega_z}$, 
where $m_{Cr}$  is the Chromium mass. 
In a calculations for other atomic species, we use $^{168}$Er with $a^{Er}_{dd}=70a_B$ and $^{164}$Dy with $a^{Dy}_{dd}= 140a_B$, we keep the same \textbf{density} as for Chromium.
Thus the geometry is changed in these cases: for atoms with mass $m$ we choose the trapping frequency $\omega_z^{\prime} = \frac{m_{Cr}}{m}\omega_z$
 whereas the two dimensional box and the length unit $l_z$ remain the same. 

% We focus on parameter space relevant to experiments with Chromium in a stable regime. 
% None of the presented curves is in the regime where a roton in the spectrum \eqref{eqn:bogol} appears.
\begin{figure}
 \includegraphics[width=6cm]{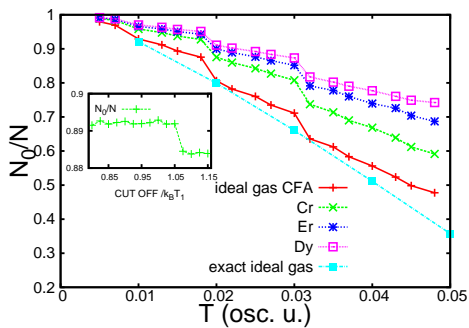}
\caption{(Color online)  Average fraction of condensed atoms. %and the third order correlation function.
 Parameters: $a = 10a_B$, "the dipolar interaction length" corresponds to different atomic species $^{52}$Cr,  $^{168}$Er,  $^{164}$Dy. 
The average density in the middle of the trap $10^{14}/\text{cm}^3$. 
Note, that oscillatory units are different for different species, as described in the text.
The exact calculation for ideal gas follows the procedure from  \cite{wilkens1997}.
In inset: fraction of condensed Cr atoms at $T_1=0.015\hbar\omega_z/k_B$ but for different values of cut-off.
\label{fig:n0}}
\end{figure}
For such parameters the fraction of atoms in the zero-momentum mode, shown in Fig. \ref{fig:n0}, is only slightly altered 
by interactions. We choose the scattering length $10a_B$, which is below "the dipolar length" $a_{dd} = 16a_B$, to enhance the 
effect of dipolar forces.
We restrict the study to temperatures up to about $100$nK. Then the cutt-off in momentum space $k_{max}$, limits also 
the theoretical spatial resolution. Here, the spatial resolution is $> 10 l_z$, which is not much smaller than the observed
structure, which could slightly impact the outcome of the calculation.

The dipolar interaction %does not change the system significantly, it only 
slows down a thermal depletion of the dominant mode.
This is mostly because of dipole repulsion - the experiment is in such a regime, that contribution of the long range part of the dipolar potential 
\eqref{eqn:edipLongRange}
remains small. Thus, the main effect of the dipolar interaction is just a shift of the contact interaction coupling constant.
It has already been proven  \cite{andersen2004}, that contact interactions in the box reduce the evaporation to higher modes, 
as we also see in Fig. \ref{fig:n0}. 
The latter can be guessed also from the Bogoliubov spectrum \eqref{eqn:bogol}: the energy gap between the $\bm{0}$ and $\bm{1}$ momentum modes
is increasing with both $g$ and $g_{dd}$.
Note the opposite behavior for the gas trapped in a harmonic potential   \cite{bisset2012, ticknor2012}. 
The latter can be qualitatively understood with the help of the local density approximation, in which
the system is considered as a set of boxes with varying chemical potential.
If the local density $n_{loc}$ in a box decreases, while temperature is constant, then the fraction: average density of the atoms in the
 $0$-momentum mode divided by $n_{loc}$, decreases also.
Due to the repulsive forces the cloud trapped in a harmonic potential widens, and in consequence in the center of the trap (so in the region
contributing mostly to the condensate) the density decreases.
Thus in the dense part of the cloud atoms evaporate to higher energy modes.
The same reasoning leads to the conclusion, that in the tails of the cloud,
where the local density would increase after the cloud broadening, the
occupation of the $\bm{0}$-momentum mode increases.
However, it turns out that globally, when repulsive forces increase, the
condensate in trap evanesces.
% 
% Thus more atoms in total is transferred to a thermal gas, and BEC
% 
% There, the repulsive forces lead to the spatial broadening of the cloud, and in consequence it to the reduction of the local density.
% , which eventually leads to a faster evaporation of the BEC.

It is worth noticing, that the repulsive dipolar force has been already used to stabilize the system against attractive contact interactions  \cite{mueller2011}
and even to suppress chemical reactions between KRb molecules  \cite{miranda2011} and dipolar relaxation for spinor gases \cite{pasquiou2010}.

The discontinuities visible in Fig. \ref{fig:n0} are due to the discretness of the finite-system spectrum. 
Precisely, there are regions where the temperature rises, and consequently cut-off energy, but the number of modes 
remains unchanged. At some value of temperature new four modes are accepted, which results in a small jump in statistical quantities.
Our qualitative results do not depend strongly on this technical problem.
Moreover in the rest of the paper we focus on such temperatures, at which the energy of the last retained mode is equal exactly to the cut-off energy.
The problem may be overcome by taking the weighted averages obtained for the "critical" cut-offs. 

In inset in Fig. \ref{fig:n0} we investigated the sensitivity on the cut-off. 
We were changing the numerical factor in Eq. \eqref{eqn:kvst} keeping constant temperature $T_1=0.015\hbar\omega_Z/k_B$. Even for a cut-off varying up to $20$\% the fraction of condensed
atoms is varying much less than $2$\%. One can see in the inset, that the statistical error is small - the main uncertainty comes from the discreetness
of the spectrum, which cause the jump at the cut-off energy $1.05k_BT$.
\begin{figure}
a)

 \includegraphics[width=6cm]{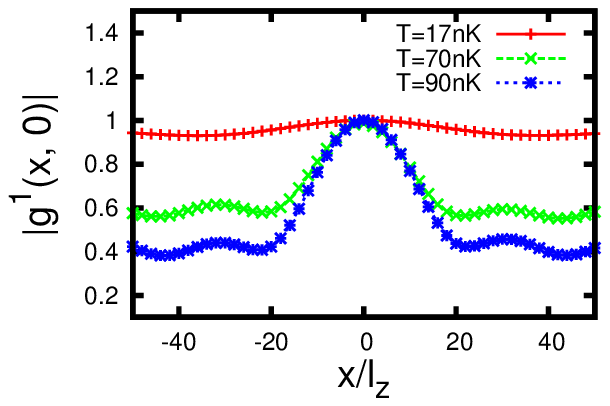}

b)

 \includegraphics[width=6cm]{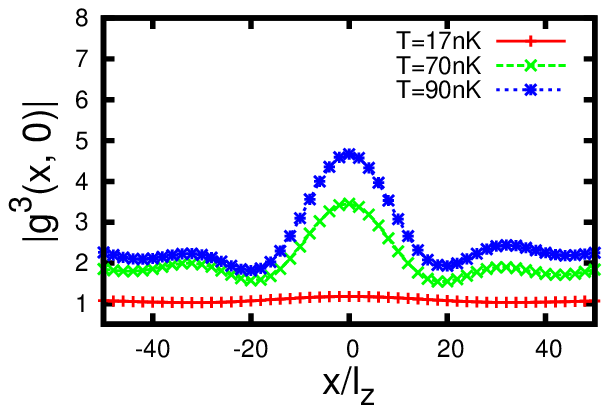}
\caption{(Color online) Cuts of the spatial first (the upper panel) and the third (the lower panel) order 
correlation functions at few chosen temperatures relevant to the experiment \cite{mueller2011}. }
\label{fig:g1g3}
\end{figure}

The most interesting quantities are correlation functions. Among them, the crucial role for identifying condensation is played by the first order 
correlation function defined here as
\begin{equation}
 g^1(\bm{\rho}) = \meanv{\Psi^* (\bm{\rho}) \Psi ( \bm{0} )}/n_{2D}.
\end{equation}
In the 2D uniform system the true condensation does not exist at $T>0$. Indeed in 
 Fig. \ref{fig:g1g3}a) one can see that even for very low temperatures  $g^1 (\bm{\rho})$ is decaying slightly below $1$ for $\bm{\rho} \neq \bm{0}$.
The width of this function defines the correlation length. 
It decreases with increasing temperature making the coherent region much smaller than the size of the system. 
In our regime, where both contact and dipolar interactions are weak (comparing to the contact interaction at the background scattering length), this reduction is mostly due to a shortening 
of the de Broglie wavelength $\lambda_d = \frac{h}{\sqrt{2\pi k_B T}}$. 
For temperatures used in Fig. \ref{fig:g1g3} the $\lambda_d$ is varies between $11l_z$ and $25l_z$.

The possible density fluctuation are quantified with the help of the third order correlation function. Here we focus on its normalized cut:
\begin{equation}
 g^3 \bb{\bm{\rho}} = \meanv{ |\Psi (\bm{\rho})|^2 |\Psi (\bm{0})|^4}/n_{2D}^3.
\end{equation}
We choose this quantity instead of $g^2$ as it has a direct impact on the rate of three body losses \cite{burt1997}, 
thus it is relevant for the condensate lifetime. 
Fig. \ref{fig:g1g3}b) shows, that surprisingly, in experiment  \cite{mueller2011} the gas was probably highly bunched, with a
 few times faster losses (at $T>100$nK) comparing with the pure condensate ($T=0K$). 
The bunching is simply due to approaching the thermal state, at which $g^3=6$ is expected.

\begin{figure}
 \includegraphics[width=6cm]{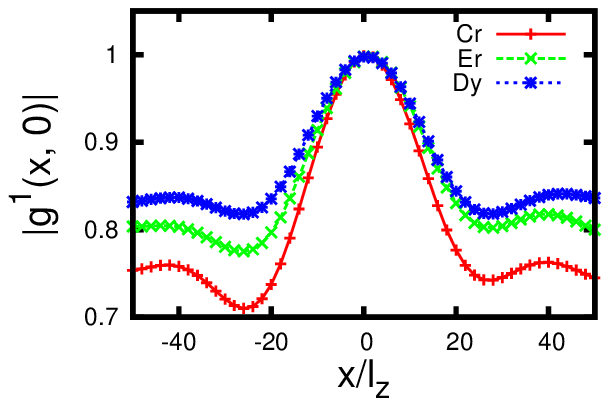}
 \includegraphics[width=6cm]{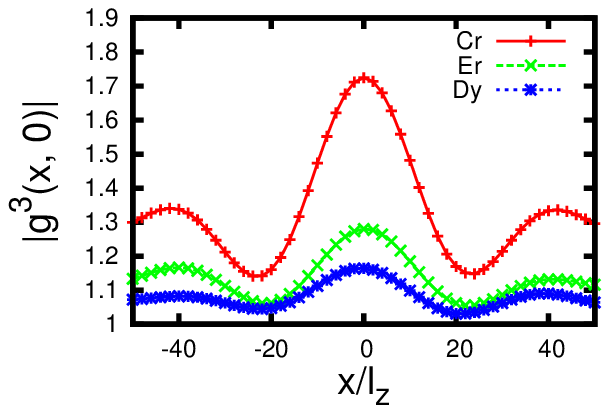}
\caption{(Color online)  The first (the upper panel)  and the third (the lower panel) order correlation functions at temperature 
$T=0.020\hbar\omega_z /k_B$ ($\approx 30$nK for Cr
%the reference experiment  \cite{mueller2011}
) for $a=10a_B$ but for different values of $a_{dd}$ corresponding to different atomic species.}
\label{fig:g1g3Comp}
\end{figure}
In Fig. \ref{fig:g1g3Comp} we compare the correlations of Chromium with the ones of Erbium and Dysprosium. 
As can be seen the correlation length is increasing with increasing dipolar forces and at the same time bunching is getting weaker
\footnote{Similar behavior has been observed for the $g^2$ function.}. 
\begin{figure}
\includegraphics[width=6cm]{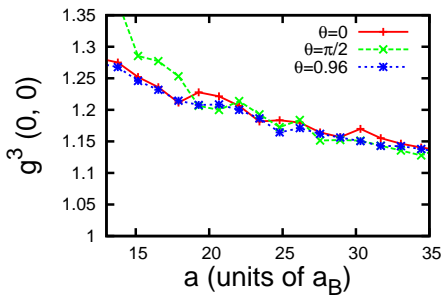}
\caption{(Color online) The third order correlation function $g^3 (0, 0)$. The role 
of dipolar interaction increases when approaching the instability border. The temperature $T=0.02\hbar\omega_z/k_B$ \label{fig:g3T}}
\end{figure}
These effects may be understood with the help of Fig. \ref{fig:g3T} where $g^3(0, 0)$ versus scattering length is shown.
%Strange results for Dysprosium are again caused by the collapse even for low temperatures.
For large scattering length the value of $g^3$ does not depend strongly on the dipolar interactions, 
as it is insensitive to different polarization angles.
This increase for smaller and smaller $a$ is probably caused by approaching unstable regions. 
The latter is suggested by the rapid bunching of the least stable configuration $\theta=\pi/2$. 
In such a geometry, when $a$ reaches some critical value in the Bogoliubov spectrum  imaginary frequencies appear, implying a phonon instability
\footnote{The stability region may grow with increasing temperature [14, 15].}.
Then, in a collapse the gas would shrink almost to a delta function. 
Thus in the unstable region the system should be a mixture of infinitely many plane waves.
%Clearly in such situation the CFA need substantial modifications.
This qualitative analysis indicates however
that also close to instabilities, (but still in a stable regime) the occupation of $\bm{k} \neq 0$ modes should already increase. 
This is the regime studied throughout the paper - where one can see already dipolar effects, although the Bogoliubov spectrum does not reach roton yet. 
The macroscopically occupied higher modes give an increase of the $g^3$ function. 
However, the dipolar interactions stabilize the system, due to their repulsive part.
One can see already in Fig. \ref{fig:n0} that the occupation of the $\bm{0}$-momentum mode increases with the increase of the dipole moment.
Thus the role of excitations is getting smaller for stronger magnets, making the atomic cloud more coherent and less bunched, 
as shown in Fig \ref{fig:g1g3Comp}. 
% Note, that the $g^1$ function is nothing else but the position-dependent reduced density matrix. 
% The reduction of coherence length investigated directly from the reduced density matrix analysis has been presented i.e. in \cite{bienias2013Proc}.
\section{Conclusions\label{sec:conclusion}}
We have presented the classical field approximation for a weakly interacting dipolar gas.
%The interesting features are changes in the stability diagram. 
%System is proven to be less stable with increasing temperature compared to results from Bogoliubov approximation.
We show the statistical properties relevant for present experiments with Chromium. 
The dipolar forces there are only a small perturbation around the ideal gas case. 
However, at the temperature at which the experiment  \cite{mueller2011} was done, 
the third order correlation function is highly increased compared to 
the zero temperature case.
Due to this bunching, the three body losses should be a few times faster.
At the same temperature the correlation length is strongly reduced. In the case of atoms with stronger dipole moment, Erbium and Dysprosium,
the reduction of the size of the coherent regions is getting smaller.
\begin{acknowledgments}
\label{sec:acknowl}
We are grateful to Stephan M\"uller for helpful discussions.
Three of us (K. P., P. B. and K. Rz.) acknowledge the support from the (Polish) National Science Center
under contract No. DEC-2011/01/B/ST2/04069.
All authors acknowledge the financial support from the project \qu{Decoherence in long range interacting
quantum systems and devices} supported by contract research "Internationale Spitzenforschung II-2" of the Baden-W\"urttemberg Stiftung.
\end{acknowledgments}

\appendix

\section{Derivation of the dipolar interaction energy\label{appendix:DipolarInteraction}}

Dipolar energy in the classical field approximation has a form:
\begin{equation}
 E_{dip} = \frac{1}{2} \intr\intrp \, |\psiC |^2  |\psiCp|^2 \vdipDif,
\label{eqn:Edip}
\end{equation}
where the three dimensional classical field
\begin{equation}
 \Psi^{3D}(\bm{r}) = \sqrt{N}\psi_0 (z) \sum_{\bm{i}}^{K_{max}} \alpha_{\bm{i} }\frac{ e^{ \ii \bm{k}_{ \bm{i}} \bm{\rho} } }{L},
\label{eqn:psiC}
\end{equation}
is a counterpart of the field operator and $\vdip$ is a dipolar interaction potential for atoms polarized in the $Z$ direction given in Eq. \eqref{eqn:vdip}.
Here $\bm{i}$ denotes a two-dimensional index $[i_x, i_y]$ and the sum in \eqref{eqn:psiC} is over all indices $\bm{i}$ which fulfil the condition $\sqrt{i_x^2 + i_y^2}\leq K_{max}$. 
The periodic boundary conditions enforce discrete values of momenta \eqref{eqn:discrete-momenta}.
The integral $\intr$ in \eqref{eqn:Edip} is over all points from  the domain 
$[-\frac{L}{2}, \frac{L}{2}] \times [-\frac{L}{2}, \frac{L}{2}] \times \mathbb{R}$, 
where functions have Fourier transform of the form
\begin{equation}
 h\bb{\bm{r}} = \frac{1}{L^2} \int \frac{\mathrm{d}\!k_z }{\sqrt{2\pi}}\sum_{\bm{i}} h_{\bm{i}} \bb{k_z} e^{\ii \ki{i}\bm{\rho} + \ii k_z z}.
\label{eqn:fourierSemi}
\end{equation}
Assuming dipoles oriented within the $XZ$-plane the expansion coefficient of the dipolar potential is equal to 
\begin{equation}
 V_{\bm{i}} \bb{k_z}= \frac{g_{dd}}{\sqrt{2\pi}} \bb{3\frac{\bb{k_{\bm{i}}^{(x)}}^2\sin^2\bb{\theta} + k_z^2\cos^2\bb{\theta}}{k_z^2 + k_{\bm{i}}^2} -1},
\label{eqn:vdipF}
\end{equation}
where $\theta$ is an angle between polarization of all dipoles and the $Z$ direction and $k_{\bm{i}}^{(x)}=\frac{2\pi}{L} [i_x, i_y]$ 
is the $x$ component of the vector $\bm{k}$ and $\bm{k}_{\bm{i}}$ is the projection of $\bm{k}$ onto $XY$ - plane.

\begin{widetext}
Combining Eqns. \eqref{eqn:vdipF}, \eqref{eqn:fourierSemi} and \eqref{eqn:psiC} in Eq. \eqref{eqn:Edip} one gets
\begin{eqnarray}
\nonumber E_{dip} &=& \frac{1}{2} \intr\intrp |\psiC |^2  |\psiCp|^2 \vdipDif =\frac{1}{2L^2}\int \frac{\mathrm{d}^3r\mathrm{d}^3r^{\prime}\mathrm{d}\!k_z }
{\sqrt{2\pi}}\sum_{\bm{i}} |\psiC |^2  |\psiCp|^2 V_{\bm{i}} \bb{k_z} e^{\ii \ki{i}\bb{\bm{\rho}-\bm{\rho}^{\prime}} + \ii k_z \bb{z-z^{\prime}}}\\
 &=& \frac{N^2}{2L^2}\int \frac{\mathrm{d}\!k_z }{\sqrt{2\pi}} \sum_{\bm{i}} \bb{\intI{z}{\psi_0^2(z) e^{\ii k_z z}}}\bb{\intI{z^\prime}{\psi_0^2(z^{\prime}) e^{-\ii k_{z^{\prime}} z^{\prime}}}} 
 \bb{\frac{1}{L^2}\intI{\bm{\rho}}{\sum_{\bm{m}, \bm{n}}^{K_{max}}\alpha_{\bm{m}}^*\alpha_{\bm{n}} e^{-\ii \bb{\bm{n} - \bm{m} - \ki{i}}\bm{\rho}}}}\\
\nonumber  &\times & \bb{\frac{1}{L^2}\intI{\bm{\rho}^{\prime}}{\sum_{\bm{p}, \bm{q}}^{K_{max}}\alpha_{\bm{p}}^*\alpha_{\bm{q}} e^{-\ii \bb{\bm{q} - \bm{p} + \ki{i}}\bm{\rho}^{\prime}}}} V_{\bm{i}} \bb{k_z} 
\label{eqn:edip2}
\end{eqnarray}
\end{widetext}
Thus the expression for the dipolar energy consists of a product of four separate integrals (related to Fourier transforms), 
 afterwards integrated over $k_z$. There are two integrals which are Fourier transforms of $\psi_0^2$:
\begin{equation}
 \intI{z}{\psi_0^2(z) e^{\pm \ii k_z z}} = e^{-k_z^2 l_z^2/4}.
\end{equation}
Both integrals over coordinates  in the box lead to the Kronecker delta functions:
\begin{equation*}
 \int \mbox{d}^2\rho {\sum_{\bm{m}, \bm{n}}^{K_{max}}\alpha_{\bm{m}}^*\alpha_{\bm{n}} e^{-\ii \bb{\bm{n} - \bm{m} \pm \ki{i}}\bm{\rho}}}=L^2\sum_{\bm{m}, \bm{n}}^{K_{max}}\alpha_{\bm{m}}^*\alpha_{\bm{n}}\delta_{\bm{n} - \bm{m}}^{\pm\ki{i}}.
\end{equation*}
%ZLE:Note, that only terms which consists of $\ki{i} < K_{max}$ gives non-zero contribution.

Thus the formula for dipolar interaction energy \eqref{eqn:Edip} reduces to:
\begin{eqnarray*}
 E_{dip} =& \frac{N^2}{2L^2}\int \frac{\mathrm{d}\!k_z }{\sqrt{2\pi}}e^{-k_z^2 l_z^2/2} \times\\
  &\sum_{\bm{i}, \bm{m}, \bm{n}, \bm{p}, \bm{q}}^{K_{max}} \alpha_{\bm{m}}^*\alpha_{\bm{p}}^*\alpha_{\bm{q}}\alpha_{\bm{n}}\delta_{\bm{n} - \bm{m}}^{\ki{i}} \delta_{\bm{q} - \bm{p}}^{-\ki{i}} V_{\bm{i}} \bb{k_z}\\
  = & \frac{n_{2D}N}{2}\int \frac{\mathrm{d}\!k_z }{\sqrt{2\pi}}e^{-k_z^2 l_z^2/2} T_{mpn} V_{\bm{n}-\bm{m}} \bb{k_z},
\end{eqnarray*}
where in the last equation two delta functions reduce sums over $\bm{i}$ and $\bm{q}$. With $n_{2D}=N/L^2$ we denote a 2D density and 
for brevity we introduce an operator:
\begin{displaymath}
 T_{mpn} = \sum_{\bm{m}, \bm{p}, \bm{n}}^{K_{max}} \alpha_{\bm{m}}^*\alpha_{\bm{p}}^*\alpha_{\bm{n}}\alpha_{\bm{m}+\bm{p}-\bm{n}} .
\end{displaymath}
Hence, the formula for dipolar interaction energy is reduced just to one integral (over $k_z$) and sum over three 2D indices ($\bm{m}, \bm{n}, \bm{p}$).
Substituting finally the explicit expression for the dipolar potential \eqref{eqn:vdipF} we have to deal with an integral
\begin{equation}
\int \mathrm{d}\!k_z\,e^{-k_z^2 l_z^2/2}
 \bb{3\frac{|k_{n_x-m_x}|^2\sin^2\theta + k_z^2\cos^2\theta}{k_z^2 + |\ki{n-m}|^2}-1}.
\label{eqn:mainInt} 
\end{equation}
The dipolar potential may be rearranged 
\begin{eqnarray*}
3\frac{k_x^2\sin^2\theta + k_z^2\cos^2\theta}{k^2}-1 &=&  \cos 2\theta + \cos^2\theta \\
 & &- 3\frac{k_x^2\cos 2\theta + k_y^2\cos^2\theta}{k^2}
\end{eqnarray*}
to a form more convenient for integral over $k_z$.
\begin{figure}
\hspace{0.5cm}(a) \hspace{3.5cm} (b)

\includegraphics[width=4cm]{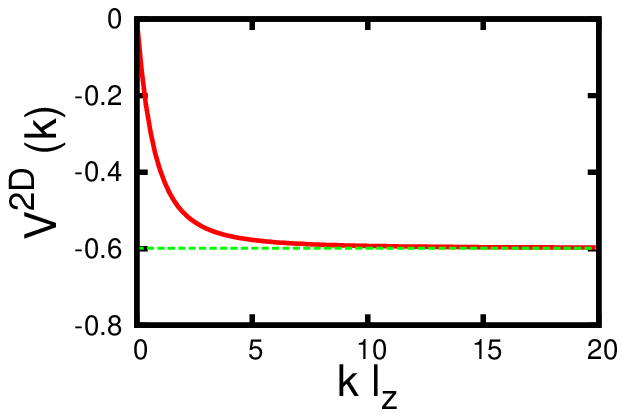}
\includegraphics[width=4cm]{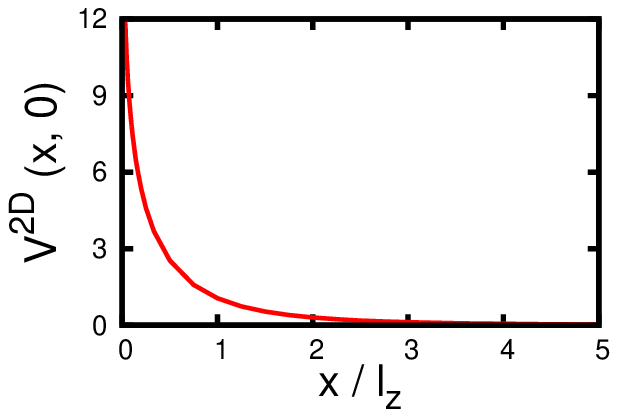}

(c)

\includegraphics[width=8cm]{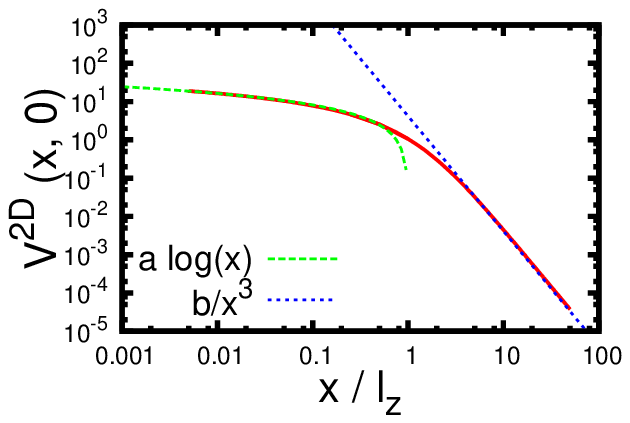}

\caption{(Color online)  a) Function $\tilde{V}^{2D}$ defined as in the equation \eqref{eqn:vdip2D} for $\theta = 0$ (polarization in $Z$ direction) in momentum space. b) Function ${V}^{2D}$ defined as in the equation \eqref{eqn:vdip2D} for $\theta = 0$ (polarization in $Z$ direction) in position and c) its scaling for small and large $\rho$ (c).}
\label{fig:vdip2D}
\end{figure}
This rearrangement allows us to write the integral over momenta in the $Z$ direction \eqref{eqn:mainInt} as a combination of two converging integrals:
\begin{displaymath}
 \int \frac{\mathrm{d}\!k_z }{\sqrt{2\pi}}e^{-k_z^2 l_z^2/2} = \frac{1}{l_z}
\end{displaymath}
and
\begin{displaymath}
 \int \frac{\mathrm{d}\!k_z }{\sqrt{2\pi}}\frac{e^{-k_z^2 l_z^2/2}}{k_z^2 + |\ki{n-m}|^2} 
= \frac{\sqrt{\pi}e^{\bb{k_{\bm{n}-\bm{m}}l_z}^2/2}}{\sqrt{2} k_{\bm{n}-\bm{m}}} \mathrm{Erfc} \frac{k_{\bm{n}-\bm{m}}l_z}{\sqrt{2}},
\end{displaymath}
where $k_{\bm{n}-\bm{m}} = \frac{2\pi}{L}\sqrt{\bb{m_x-n_x}^2 + \bb{m_y-n_y}^2}$ is just a length of the difference $\ki{m}$ and $\ki{n}$.
Finally the dipolar energy is a sum of two terms
\begin{itemize}
 \item[I] Part of dipolar potential which mimics the contact interactions:
\begin{equation} 
E_{dip}^I =   \frac{g_{dd}n_{2D}N \bb{\cos 2\theta + \cos^2\theta}}{2\sqrt{2\pi}l_z}T_{mpn} .
\end{equation}
\item[II] Part of dipolar potential which depends on momenta in a non-trivial way:
\begin{equation}
E_{dip}^{II} =   \frac{g_{dd}n_{2D}N}{l_z}T_{mpn} \tilde{V}^{2D}(\ki{n}-\ki{m}),
\label{eqn:longRangePart}
\end{equation}
with
\begin{equation}
\tilde{V}^{2D}(\bm{k}) =-\frac{3l_z}{4} \frac{k_x^2\cos 2\theta + k_y^2\cos^2\theta}{k}  e^{k^2 l_z^2/2} \mathrm{Erfc} \frac{kl_z}{\sqrt{2}} .
\label{eqn:vdip2D}
\end{equation}
\end{itemize}
We illustrate the features of the potential in a particular case of $\theta=0$.
As shown in Fig. \ref{fig:vdip2D}a) in this case the function \eqref{eqn:vdip2D} for $k\to \infty$ converges to a non-zero value.
In the position representation this asymptotic behavior leads to an additional delta function, precisely:
\begin{equation}
\mathcal{F}\bb{\tilde{V}^{2D}(\bm{k}) }= -\frac{3}{4}\sqrt{\frac{2}{\pi}}\delta \bb{\rho} + V^{2D} (\bm{\rho}),
\label{eqn:vdip2D2}
\end{equation}
where the potential in a position representation is given by
\begin{equation}
 V^{2D} (\bm{\rho})=2\pi\int\!\mathrm{d}k\;J_0\bb{k \rho} \bb{\tilde{V}^{2D}\bb{k}+\frac{3}{4}\sqrt{\frac{2}{\pi}}} k,
\end{equation}
where $J_0\bb{x}$ is a zero order Bessel function.
The shape of this potential is plotted in Fig. \ref{fig:vdip2D}b). It diverges logarithmically at origin and tends to $0$ like $\frac{1}{\rho^3}$ for large $\rho$, as shown in Fig. \ref{fig:vdip2D}c).


\begin{thebibliography}{10}%
\makeatletter
\providecommand \@ifxundefined [1]{%
 \ifx #1\undefined \expandafter \@firstoftwo
 \else \expandafter \@secondoftwo
\fi
}%
\providecommand \@ifnum [1]{%
 \ifnum #1\expandafter \@firstoftwo
 \else \expandafter \@secondoftwo
\fi
}%
\providecommand \enquote [1]{``#1''}%
\providecommand \bibnamefont  [1]{#1}%
\providecommand \bibfnamefont [1]{#1}%
\providecommand \citenamefont [1]{#1}%
\providecommand\href[0]{\@sanitize\@href}%
\providecommand\@href[1]{\endgroup\@@startlink{#1}\endgroup\@@href}%
\providecommand\@@href[1]{#1\@@endlink}%
\providecommand \@sanitize [0]{\begingroup\catcode`\&12\catcode`\#12\relax}%
\@ifxundefined \pdfoutput {\@firstoftwo}{%
 \@ifnum{\z@=\pdfoutput}{\@firstoftwo}{\@secondoftwo}%
}{%
 \providecommand\@@startlink[1]{\leavevmode}%
 \providecommand\@@endlink[0]{}%
}{%
 \providecommand\@@startlink[1]{%
  \leavevmode
  \pdfstartlink
   attr{/Border[0 0 1 ]/H/I/C[0 1 1]}%
   user{/Subtype/Link/A<</Type/Action/S/URI/URI(#1)>>}%
  \relax
 }%
 \providecommand\@@endlink[0]{\pdfendlink}%
}%
\providecommand \url  [0]{\begingroup\@sanitize \@url }%
\providecommand \@url [1]{\endgroup\@href {#1}{\urlprefix}}%
\providecommand \urlprefix [0]{URL }%
\providecommand \Eprint[0]{\href }%
\@ifxundefined \urlstyle {%
  \providecommand \doi [1]{doi:\discretionary{}{}{}#1}%
}{%
  \providecommand \doi [0]{doi:\discretionary{}{}{}\begingroup
  \urlstyle{rm}\Url }%
}%
\providecommand \doibase [0]{http://dx.doi.org/}%
\providecommand \Doi[1]{\href{\doibase#1}}%
\providecommand \bibAnnote [3]{%
  \BibitemShut{#1}%
  \begin{quotation}\noindent
    \textsc{Key:}\ #2\\\textsc{Annotation:}\ #3%
  \end{quotation}%
}%
\providecommand \bibAnnoteFile [2]{%
  \IfFileExists{#2}{\bibAnnote {#1} {#2} {\input{#2}}}{}%
}%
\providecommand \typeout [0]{\immediate \write \m@ne }%
\providecommand \selectlanguage [0]{\@gobble}%
\providecommand \bibinfo [0]{\@secondoftwo}%
\providecommand \bibfield [0]{\@secondoftwo}%
\providecommand \translation [1]{[#1]}%
\providecommand \BibitemOpen[0]{}%
\providecommand \bibitemStop [0]{}%
\providecommand \bibitemNoStop [0]{.\EOS\space}%
\providecommand \EOS [0]{\spacefactor3000\relax}%
\providecommand \BibitemShut [1]{\csname bibitem#1\endcsname}%
%</preamble>
\bibitem{mueller2011}%
  \BibitemOpen
  \bibfield{author}{%
  \bibinfo {author} {\bibfnamefont{S.}~\bibnamefont{M\"uller}}, \bibinfo
  {author} {\bibfnamefont{J.}~\bibnamefont{Billy}}, \bibinfo {author}
  {\bibfnamefont{E.~A.~L.}\ \bibnamefont{Henn}}, \bibinfo {author}
  {\bibfnamefont{H.}~\bibnamefont{Kadau}}, \bibinfo {author}
  {\bibfnamefont{A.}~\bibnamefont{Griesmaier}}, \bibinfo {author}
  {\bibfnamefont{M.}~\bibnamefont{Jona-Lasinio}}, \bibinfo {author}
  {\bibfnamefont{L.}~\bibnamefont{Santos}},\ and\ \bibinfo {author}
  {\bibfnamefont{T.}~\bibnamefont{Pfau}},\ }%
  \bibfield{journal}{%
  \Doi{10.1103/PhysRevA.84.053601}{\bibinfo {journal} {Phys. Rev. A}}\ }%
  \textbf{\bibinfo {volume} {84}},\ \bibinfo {pages} {053601} (\bibinfo {month}
  {Nov}\ \bibinfo {year} {2011}),\
  \url{http://link.aps.org/doi/10.1103/PhysRevA.84.053601}%
  \bibAnnoteFile{NoStop}{mueller2011}%
\bibitem{griesmaier2005}%
  \BibitemOpen
  \bibfield{author}{%
  \bibinfo {author} {\bibfnamefont{A.}~\bibnamefont{Griesmaier}}, \bibinfo
  {author} {\bibfnamefont{J.}~\bibnamefont{Werner}}, \bibinfo {author}
  {\bibfnamefont{S.}~\bibnamefont{Hensler}}, \bibinfo {author}
  {\bibfnamefont{J.}~\bibnamefont{Stuhler}},\ and\ \bibinfo {author}
  {\bibfnamefont{T.}~\bibnamefont{Pfau}},\ }%
  \bibfield{journal}{%
  \Doi{10.1103/PhysRevLett.94.160401}{\bibinfo {journal} {Phys. Rev. Lett.}}\
  }%
  \textbf{\bibinfo {volume} {94}},\ \bibinfo {pages} {160401} (\bibinfo {month}
  {Apr}\ \bibinfo {year} {2005}),\
  \url{http://link.aps.org/doi/10.1103/PhysRevLett.94.160401}%
  \bibAnnoteFile{NoStop}{griesmaier2005}%
\bibitem{lahaye2008}%
  \BibitemOpen
  \bibfield{author}{%
  \bibinfo {author} {\bibfnamefont{T.}~\bibnamefont{Lahaye}}, \bibinfo {author}
  {\bibfnamefont{J.}~\bibnamefont{Metz}}, \bibinfo {author}
  {\bibfnamefont{B.}~\bibnamefont{Fr\"ohlich}}, \bibinfo {author}
  {\bibfnamefont{T.}~\bibnamefont{Koch}}, \bibinfo {author}
  {\bibfnamefont{M.}~\bibnamefont{Meister}}, \bibinfo {author}
  {\bibfnamefont{A.}~\bibnamefont{Griesmaier}}, \bibinfo {author}
  {\bibfnamefont{T.}~\bibnamefont{Pfau}}, \bibinfo {author}
  {\bibfnamefont{H.}~\bibnamefont{Saito}}, \bibinfo {author}
  {\bibfnamefont{Y.}~\bibnamefont{Kawaguchi}},\ and\ \bibinfo {author}
  {\bibfnamefont{M.}~\bibnamefont{Ueda}},\ }%
  \bibfield{journal}{%
  \Doi{10.1103/PhysRevLett.101.080401}{\bibinfo {journal} {Phys. Rev. Lett.}}\
  }%
  \textbf{\bibinfo {volume} {101}},\ \bibinfo {pages} {080401} (\bibinfo
  {month} {Aug}\ \bibinfo {year} {2008}),\
  \url{http://link.aps.org/doi/10.1103/PhysRevLett.101.080401}%
  \bibAnnoteFile{NoStop}{lahaye2008}%
\bibitem{goral2002}%
  \BibitemOpen
  \bibfield{author}{%
  \bibinfo {author} {\bibfnamefont{K.}~\bibnamefont{G\'oral}}, \bibinfo
  {author} {\bibfnamefont{L.}~\bibnamefont{Santos}},\ and\ \bibinfo {author}
  {\bibfnamefont{M.}~\bibnamefont{Lewenstein}},\ }%
  \bibfield{journal}{%
  \Doi{10.1103/PhysRevLett.88.170406}{\bibinfo {journal} {Phys. Rev. Lett.}}\
  }%
  \textbf{\bibinfo {volume} {88}},\ \bibinfo {pages} {170406} (\bibinfo {month}
  {Apr}\ \bibinfo {year} {2002}),\
  \url{http://link.aps.org/doi/10.1103/PhysRevLett.88.170406}%
  \bibAnnoteFile{NoStop}{goral2002}%
\bibitem{santos2003}%
  \BibitemOpen
  \bibfield{author}{%
  \bibinfo {author} {\bibfnamefont{L.}~\bibnamefont{Santos}}, \bibinfo {author}
  {\bibfnamefont{G.~V.}\ \bibnamefont{Shlyapnikov}},\ and\ \bibinfo {author}
  {\bibfnamefont{M.}~\bibnamefont{Lewenstein}},\ }%
  \bibfield{journal}{%
  \Doi{10.1103/PhysRevLett.90.250403}{\bibinfo {journal} {Phys. Rev. Lett.}}\
  }%
  \textbf{\bibinfo {volume} {90}},\ \bibinfo {pages} {250403} (\bibinfo {month}
  {Jun}\ \bibinfo {year} {2003}),\
  \url{http://link.aps.org/doi/10.1103/PhysRevLett.90.250403}%
  \bibAnnoteFile{NoStop}{santos2003}%
\bibitem{aikawa2012}%
  \BibitemOpen
  \bibfield{author}{%
  \bibinfo {author} {\bibfnamefont{K.}~\bibnamefont{Aikawa}}, \bibinfo {author}
  {\bibfnamefont{A.}~\bibnamefont{Frisch}}, \bibinfo {author}
  {\bibfnamefont{M.}~\bibnamefont{Mark}}, \bibinfo {author}
  {\bibfnamefont{S.}~\bibnamefont{Baier}}, \bibinfo {author}
  {\bibfnamefont{A.}~\bibnamefont{Rietzler}}, \bibinfo {author}
  {\bibfnamefont{R.}~\bibnamefont{Grimm}},\ and\ \bibinfo {author}
  {\bibfnamefont{F.}~\bibnamefont{Ferlaino}},\ }%
  \bibfield{journal}{%
  \Doi{10.1103/PhysRevLett.108.210401}{\bibinfo {journal} {Phys. Rev. Lett.}}\
  }%
  \textbf{\bibinfo {volume} {108}},\ \bibinfo {pages} {210401} (\bibinfo
  {month} {May}\ \bibinfo {year} {2012}),\
  \url{http://link.aps.org/doi/10.1103/PhysRevLett.108.210401}%
  \bibAnnoteFile{NoStop}{aikawa2012}%
\bibitem{Lu2011}%
  \BibitemOpen
  \bibfield{author}{%
  \bibinfo {author} {\bibfnamefont{M.}~\bibnamefont{Lu}}, \bibinfo {author}
  {\bibfnamefont{N.~Q.}\ \bibnamefont{Burdick}}, \bibinfo {author}
  {\bibfnamefont{S.~H.}\ \bibnamefont{Youn}},\ and\ \bibinfo {author}
  {\bibfnamefont{B.~L.}\ \bibnamefont{Lev}},\ }%
  \bibfield{journal}{%
  \Doi{10.1103/PhysRevLett.107.190401}{\bibinfo {journal} {Phys. Rev. Lett.}}\
  }%
  \textbf{\bibinfo {volume} {107}},\ \bibinfo {pages} {190401} (\bibinfo
  {month} {Oct}\ \bibinfo {year} {2011}),\
  \url{http://link.aps.org/doi/10.1103/PhysRevLett.107.190401}%
  \bibAnnoteFile{NoStop}{Lu2011}%
\bibitem{aikawa2010}%
  \BibitemOpen
  \bibfield{author}{%
  \bibinfo {author} {\bibfnamefont{K.}~\bibnamefont{Aikawa}}, \bibinfo {author}
  {\bibfnamefont{D.}~\bibnamefont{Akamatsu}}, \bibinfo {author}
  {\bibfnamefont{M.}~\bibnamefont{Hayashi}}, \bibinfo {author}
  {\bibfnamefont{K.}~\bibnamefont{Oasa}}, \bibinfo {author}
  {\bibfnamefont{J.}~\bibnamefont{Kobayashi}}, \bibinfo {author}
  {\bibfnamefont{P.}~\bibnamefont{Naidon}}, \bibinfo {author}
  {\bibfnamefont{T.}~\bibnamefont{Kishimoto}}, \bibinfo {author}
  {\bibfnamefont{M.}~\bibnamefont{Ueda}},\ and\ \bibinfo {author}
  {\bibfnamefont{S.}~\bibnamefont{Inouye}},\ }%
  \bibfield{journal}{%
  \Doi{10.1103/PhysRevLett.105.203001}{\bibinfo {journal} {Phys. Rev. Lett.}}\
  }%
  \textbf{\bibinfo {volume} {105}},\ \bibinfo {pages} {203001} (\bibinfo
  {month} {Nov}\ \bibinfo {year} {2010}),\
  \url{http://link.aps.org/doi/10.1103/PhysRevLett.105.203001}%
  \bibAnnoteFile{NoStop}{aikawa2010}%
\bibitem{Ni2008}%
  \BibitemOpen
  \bibfield{author}{%
  \bibinfo {author} {\bibfnamefont{K.-K.}\ \bibnamefont{Ni}}, \bibinfo {author}
  {\bibfnamefont{S.}~\bibnamefont{Ospelkaus}}, \bibinfo {author}
  {\bibfnamefont{M.~H.~G.}\ \bibnamefont{de~Miranda}}, \bibinfo {author}
  {\bibfnamefont{A.}~\bibnamefont{Pe'er}}, \bibinfo {author}
  {\bibfnamefont{B.}~\bibnamefont{Neyenhuis}}, \bibinfo {author}
  {\bibfnamefont{J.~J.}\ \bibnamefont{Zirbel}}, \bibinfo {author}
  {\bibfnamefont{S.}~\bibnamefont{Kotochigova}}, \bibinfo {author}
  {\bibfnamefont{P.~S.}\ \bibnamefont{Julienne}}, \bibinfo {author}
  {\bibfnamefont{D.~S.}\ \bibnamefont{Jin}},\ and\ \bibinfo {author}
  {\bibfnamefont{J.}~\bibnamefont{Ye}},\ }%
  \bibfield{journal}{%
  \Doi{10.1126/science.1163861}{\bibinfo {journal} {Science}}\ }%
  \textbf{\bibinfo {volume} {322}},\ \bibinfo {pages} {231} (\bibinfo {year}
  {2008}),\
  \Eprint{http://arxiv.org/abs/http://www.sciencemag.org/content/322/5899/231.%
full.pdf}{http://www.sciencemag.org/content/322/5899/231.full.pdf},\
  \url{http://www.sciencemag.org/content/322/5899/231.abstract}%
  \bibAnnoteFile{NoStop}{Ni2008}%
\bibitem{ronen2007}%
  \BibitemOpen
  \bibfield{author}{%
  \bibinfo {author} {\bibfnamefont{S.}~\bibnamefont{Ronen}}\ and\ \bibinfo
  {author} {\bibfnamefont{J.~L.}\ \bibnamefont{Bohn}},\ }%
  \bibfield{journal}{%
  \Doi{10.1103/PhysRevA.76.043607}{\bibinfo {journal} {Phys. Rev. A}}\ }%
  \textbf{\bibinfo {volume} {76}},\ \bibinfo {pages} {043607} (\bibinfo {month}
  {Oct}\ \bibinfo {year} {2007}),\
  \url{http://link.aps.org/doi/10.1103/PhysRevA.76.043607}%
  \bibAnnoteFile{NoStop}{ronen2007}%
\bibitem{hadzibabic2006}%
  \BibitemOpen
  \bibfield{author}{%
  \bibinfo {author} {\bibfnamefont{Z.}~\bibnamefont{Hadzibabic}}, \bibinfo
  {author} {\bibfnamefont{P.}~\bibnamefont{Kr\"{u}ger}}, \bibinfo {author}
  {\bibfnamefont{M.}~\bibnamefont{Cheneau}}, \bibinfo {author}
  {\bibfnamefont{B.}~\bibnamefont{Battelier}},\ and\ \bibinfo {author}
  {\bibfnamefont{J.}~\bibnamefont{Dalibard}},\ }%
  \bibfield{journal}{%
  \Doi{10.1038/nature04851}{\bibinfo {journal} {Nature}}\ }%
  \textbf{\bibinfo {volume} {441}},\ \bibinfo {pages} {1118} (\bibinfo {month}
  {jun}\ \bibinfo {year} {2006}),\ ISSN \bibinfo {issn} {1476-4687},\
  \url{http://www.ncbi.nlm.nih.gov/pubmed/16810249}%
  \bibAnnoteFile{NoStop}{hadzibabic2006}%
\bibitem{ticknor2012}%
  \BibitemOpen
  \bibfield{author}{%
  \bibinfo {author} {\bibfnamefont{C.}~\bibnamefont{Ticknor}},\ }%
  \bibfield{journal}{%
  \Doi{10.1103/PhysRevA.85.033629}{\bibinfo {journal} {Phys. Rev. A}}\ }%
  \textbf{\bibinfo {volume} {85}},\ \bibinfo {pages} {033629} (\bibinfo {month}
  {Mar}\ \bibinfo {year} {2012}),\
  \url{http://link.aps.org/doi/10.1103/PhysRevA.85.033629}%
  \bibAnnoteFile{NoStop}{ticknor2012}%
\bibitem{ticknor2012B}%
  \BibitemOpen
  \bibfield{author}{%
  \bibinfo {author} {\bibfnamefont{C.}~\bibnamefont{Ticknor}},\ }%
  \bibfield{journal}{%
  \Doi{10.1103/PhysRevA.86.053602}{\bibinfo {journal} {Phys. Rev. A}}\ }%
  \textbf{\bibinfo {volume} {86}},\ \bibinfo {pages} {053602} (\bibinfo {month}
  {Nov}\ \bibinfo {year} {2012}),\
  \url{http://link.aps.org/doi/10.1103/PhysRevA.86.053602}%
  \bibAnnoteFile{NoStop}{ticknor2012B}%
\bibitem{bisset2012}%
  \BibitemOpen
  \bibfield{author}{%
  \bibinfo {author} {\bibfnamefont{R.~N.}\ \bibnamefont{Bisset}}, \bibinfo
  {author} {\bibfnamefont{D.}~\bibnamefont{Baillie}},\ and\ \bibinfo {author}
  {\bibfnamefont{P.~B.}\ \bibnamefont{Blakie}},\ }%
  \bibfield{journal}{%
  \Doi{10.1103/PhysRevA.86.033609}{\bibinfo {journal} {Phys. Rev. A}}\ }%
  \textbf{\bibinfo {volume} {86}},\ \bibinfo {pages} {033609} (\bibinfo {month}
  {Sep}\ \bibinfo {year} {2012}),\
  \url{http://link.aps.org/doi/10.1103/PhysRevA.86.033609}%
  \bibAnnoteFile{NoStop}{bisset2012}%
\bibitem{bisset2011}%
  \BibitemOpen
  \bibfield{author}{%
  \bibinfo {author} {\bibfnamefont{R.~N.}\ \bibnamefont{Bisset}}, \bibinfo
  {author} {\bibfnamefont{D.}~\bibnamefont{Baillie}},\ and\ \bibinfo {author}
  {\bibfnamefont{P.~B.}\ \bibnamefont{Blakie}},\ }%
  \bibfield{journal}{%
  \Doi{10.1103/PhysRevA.83.061602}{\bibinfo {journal} {Phys. Rev. A}}\ }%
  \textbf{\bibinfo {volume} {83}},\ \bibinfo {pages} {061602} (\bibinfo {month}
  {Jun}\ \bibinfo {year} {2011}),\
  \url{http://link.aps.org/doi/10.1103/PhysRevA.83.061602}%
  \bibAnnoteFile{NoStop}{bisset2011}%
\bibitem{blakie2009}%
  \BibitemOpen
  \bibfield{author}{%
  \bibinfo {author} {\bibfnamefont{P.~B.}\ \bibnamefont{Blakie}}, \bibinfo
  {author} {\bibfnamefont{C.}~\bibnamefont{Ticknor}}, \bibinfo {author}
  {\bibfnamefont{A.~S.}\ \bibnamefont{Bradley}}, \bibinfo {author}
  {\bibfnamefont{A.~M.}\ \bibnamefont{Martin}}, \bibinfo {author}
  {\bibfnamefont{M.~J.}\ \bibnamefont{Davis}},\ and\ \bibinfo {author}
  {\bibfnamefont{Y.}~\bibnamefont{Kawaguchi}},\ }%
  \bibfield{journal}{%
  \Doi{10.1103/PhysRevE.80.016703}{\bibinfo {journal} {Phys. Rev. E}}\ }%
  \textbf{\bibinfo {volume} {80}},\ \bibinfo {pages} {016703} (\bibinfo {month}
  {Jul}\ \bibinfo {year} {2009}),\
  \url{http://link.aps.org/doi/10.1103/PhysRevE.80.016703}%
  \bibAnnoteFile{NoStop}{blakie2009}%
\bibitem{witkowska2009}%
  \BibitemOpen
  \bibfield{author}{%
  \bibinfo {author} {\bibfnamefont{E.}~\bibnamefont{Witkowska}}, \bibinfo
  {author} {\bibfnamefont{M.}~\bibnamefont{Gajda}},\ and\ \bibinfo {author}
  {\bibfnamefont{K.}~\bibnamefont{Rzazewski}},\ }%
  \bibfield{journal}{%
  \Doi{10.1103/PhysRevA.79.033631}{\bibinfo {journal} {Phys. Rev. A}}\ }%
  \textbf{\bibinfo {volume} {79}},\ \bibinfo {pages} {033631} (\bibinfo {month}
  {Mar}\ \bibinfo {year} {2009})%
  \bibAnnoteFile{NoStop}{witkowska2009}%
\bibitem{witkowska2010}%
  \BibitemOpen
  \bibfield{author}{%
  \bibinfo {author} {\bibfnamefont{E.}~\bibnamefont{Witkowska}}, \bibinfo
  {author} {\bibfnamefont{M.}~\bibnamefont{Gajda}},\ and\ \bibinfo {author}
  {\bibfnamefont{K.}~\bibnamefont{Rzazewski}},\ }%
  \bibfield{journal}{%
  \bibinfo {journal} {Optics {C}ommunications}\ }%
  \textbf{\bibinfo {volume} {283}},\ \bibinfo {pages} {671} (\bibinfo {year}
  {2010})%
  \bibAnnoteFile{NoStop}{witkowska2010}%
\bibitem{bienias2011}%
  \BibitemOpen
  \bibfield{author}{%
  \bibinfo {author} {\bibfnamefont{P.}~\bibnamefont{Bienias}}, \bibinfo
  {author} {\bibfnamefont{K.}~\bibnamefont{Paw\l{}owski}}, \bibinfo {author}
  {\bibfnamefont{M.}~\bibnamefont{Gajda}},\ and\ \bibinfo {author}
  {\bibfnamefont{K.}~\bibnamefont{Rzazewski}},\ }%
  \bibfield{journal}{%
  \Doi{10.1103/PhysRevA.83.033610}{\bibinfo {journal} {Phys. Rev. A}}\ }%
  \textbf{\bibinfo {volume} {83}},\ \bibinfo {pages} {033610} (\bibinfo {month}
  {Mar}\ \bibinfo {year} {2011}),\
  \url{http://link.aps.org/doi/10.1103/PhysRevA.83.033610}%
  \bibAnnoteFile{NoStop}{bienias2011}%
\bibitem{karpiuk2012}%
  \BibitemOpen
  \bibfield{author}{%
  \bibinfo {author} {\bibfnamefont{T.}~\bibnamefont{Karpiuk}}, \bibinfo
  {author} {\bibfnamefont{P.}~\bibnamefont{Deuar}}, \bibinfo {author}
  {\bibfnamefont{P.}~\bibnamefont{Bienias}}, \bibinfo {author}
  {\bibfnamefont{E.}~\bibnamefont{Witkowska}}, \bibinfo {author}
  {\bibfnamefont{K.}~\bibnamefont{Paw\l{}owski}}, \bibinfo {author}
  {\bibfnamefont{M.}~\bibnamefont{Gajda}}, \bibinfo {author}
  {\bibfnamefont{K.}~\bibnamefont{Rzazewski}},\ and\ \bibinfo {author}
  {\bibfnamefont{M.}~\bibnamefont{Brewczyk}},\ }%
  \bibfield{journal}{%
  \Doi{10.1103/PhysRevLett.109.205302}{\bibinfo {journal} {Phys. Rev. Lett.}}\
  }%
  \textbf{\bibinfo {volume} {109}},\ \bibinfo {pages} {205302} (\bibinfo
  {month} {Nov}\ \bibinfo {year} {2012}),\
  \url{http://link.aps.org/doi/10.1103/PhysRevLett.109.205302}%
  \bibAnnoteFile{NoStop}{karpiuk2012}%
\bibitem{bienias2011a}%
  \BibitemOpen
  \bibfield{author}{%
  \bibinfo {author} {\bibfnamefont{P.}~\bibnamefont{Bienias}}, \bibinfo
  {author} {\bibfnamefont{K.}~\bibnamefont{Paw\l{}owski}}, \bibinfo {author}
  {\bibfnamefont{M.}~\bibnamefont{Gajda}},\ and\ \bibinfo {author}
  {\bibfnamefont{K.}~\bibnamefont{Rzazewski}},\ }%
  \bibfield{journal}{%
  \bibinfo {journal} {EPL (Europhysics Letters)}\ }%
  \textbf{\bibinfo {volume} {96}},\ \bibinfo {pages} {10011} (\bibinfo {year}
  {2011}),\ \url{http://stacks.iop.org/0295-5075/96/i=1/a=10011}%
  \bibAnnoteFile{NoStop}{bienias2011a}%
\bibitem{wilkens1997}%
  \BibitemOpen
  \bibfield{author}{%
  \bibinfo {author} {\bibfnamefont{M.}~\bibnamefont{Wilkens}}\ and\ \bibinfo
  {author} {\bibfnamefont{C.}~\bibnamefont{Weiss}},\ }%
  \bibfield{journal}{%
  \Doi{10.1080/09500349708231847}{\bibinfo {journal} {Journal of Modern
  Optics}}\ }%
  \textbf{\bibinfo {volume} {44}},\ \bibinfo {pages} {1801} (\bibinfo {year}
  {1997}),\
  \Eprint{http://arxiv.org/abs/http://www.tandfonline.com/doi/pdf/10.1080/0950%
0349708231847}{http://www.tandfonline.com/doi/pdf/10.1080/09500349708231847},\
  \url{http://www.tandfonline.com/doi/abs/10.1080/09500349708231847}%
  \bibAnnoteFile{NoStop}{wilkens1997}%
\bibitem{metropolis1953}%
  \BibitemOpen
  \bibfield{author}{%
  \bibinfo {author} {\bibfnamefont{N.}~\bibnamefont{Metropolis}}, \bibinfo
  {author} {\bibfnamefont{A.~W.}\ \bibnamefont{Rosenbluth}}, \bibinfo {author}
  {\bibfnamefont{M.~N.}\ \bibnamefont{Rosenbluth}}, \bibinfo {author}
  {\bibfnamefont{A.~H.}\ \bibnamefont{Teller}},\ and\ \bibinfo {author}
  {\bibfnamefont{E.}~\bibnamefont{Teller}},\ }%
  \bibfield{journal}{%
  \bibinfo {journal} {J. Chem. Phys.}\ }%
  \textbf{\bibinfo {volume} {21}},\ \bibinfo {pages} {1087} (\bibinfo {year}
  {1953})%
  \bibAnnoteFile{NoStop}{metropolis1953}%
\bibitem{penrose1956}%
  \BibitemOpen
  \bibfield{author}{%
  \bibinfo {author} {\bibfnamefont{O.}~\bibnamefont{Penrose}}\ and\ \bibinfo
  {author} {\bibfnamefont{L.}~\bibnamefont{Onsager}},\ }%
  \bibfield{journal}{%
  \bibinfo {journal} {Phys. Rev. Lett.}\ }%
  \textbf{\bibinfo {volume} {104}},\ \bibinfo {pages} {576} (\bibinfo {year}
  {1956})%
  \bibAnnoteFile{NoStop}{penrose1956}%
\bibitem{blakie2012}%
  \BibitemOpen
  \bibfield{author}{%
  \bibinfo {author} {\bibfnamefont{P.~B.}\ \bibnamefont{Blakie}}, \bibinfo
  {author} {\bibfnamefont{D.}~\bibnamefont{Baillie}},\ and\ \bibinfo {author}
  {\bibfnamefont{R.~N.}\ \bibnamefont{Bisset}},\ }%
  \bibfield{journal}{%
  \Doi{10.1103/PhysRevA.86.021604}{\bibinfo {journal} {Phys. Rev. A}}\ }%
  \textbf{\bibinfo {volume} {86}},\ \bibinfo {pages} {021604} (\bibinfo {month}
  {Aug}\ \bibinfo {year} {2012}),\
  \url{http://link.aps.org/doi/10.1103/PhysRevA.86.021604}%
  \bibAnnoteFile{NoStop}{blakie2012}%
\bibitem{sykes2012}%
  \BibitemOpen
  \bibfield{author}{%
  \bibinfo {author} {\bibfnamefont{A.~G.}\ \bibnamefont{Sykes}}\ and\ \bibinfo
  {author} {\bibfnamefont{C.}~\bibnamefont{Ticknor}},\ }%
  \enquote{\bibinfo {title} {Coherence and correlation functions of quasi-2d
  dipolar superfluids at zero temperature},}\  (\bibinfo {year} {2012}),\
  \Eprint{http://arxiv.org/abs/arXiv:1206.1350}{arXiv:1206.1350}%
  \bibAnnoteFile{NoStop}{sykes2012}%
\bibitem{koch2008}%
  \BibitemOpen
  \bibfield{author}{%
  \bibinfo {author} {\bibfnamefont{T.}~\bibnamefont{Koch}}, \bibinfo {author}
  {\bibfnamefont{T.}~\bibnamefont{Lahaye}}, \bibinfo {author}
  {\bibfnamefont{J.}~\bibnamefont{Metz}}, \bibinfo {author}
  {\bibfnamefont{B.}~\bibnamefont{Fr\"ohlich}}, \bibinfo {author}
  {\bibnamefont{M.}}, \bibinfo {author}
  {\bibfnamefont{A.}~\bibnamefont{Griesmaier}},\ and\ \bibinfo {author}
  {\bibnamefont{Pfau}},\ }%
  \bibfield{journal}{%
  \Doi{10.1038/nphys887}{\bibinfo {journal} {Nature Physics}}\ }%
  \textbf{\bibinfo {volume} {4}},\ \bibinfo {pages} {218} (\bibinfo {year}
  {2008})%
  \bibAnnoteFile{NoStop}{koch2008}%
\bibitem{andersen2004}%
  \BibitemOpen
  \bibfield{author}{%
  \bibinfo {author} {\bibfnamefont{J.~O.}\ \bibnamefont{Andersen}},\ }%
  \bibfield{journal}{%
  \Doi{10.1103/RevModPhys.76.599}{\bibinfo {journal} {Rev. Mod. Phys.}}\ }%
  \textbf{\bibinfo {volume} {76}},\ \bibinfo {pages} {599} (\bibinfo {month}
  {Jul}\ \bibinfo {year} {2004}),\
  \url{http://link.aps.org/doi/10.1103/RevModPhys.76.599}%
  \bibAnnoteFile{NoStop}{andersen2004}%
\bibitem{miranda2011}%
  \BibitemOpen
  \bibfield{author}{%
  \bibinfo {author} {\bibfnamefont{M.~H.~G.}\ \bibnamefont{Miranda}}, \bibinfo
  {author} {\bibfnamefont{A.}~\bibnamefont{Chotia}}, \bibinfo {author}
  {\bibfnamefont{B.}~\bibnamefont{Neyenhuis}}, \bibinfo {author}
  {\bibfnamefont{D.}~\bibnamefont{Wang}}, \bibinfo {author}
  {\bibfnamefont{G.}~\bibnamefont{Quemener}}, \bibinfo {author}
  {\bibfnamefont{S.}~\bibnamefont{Ospelkaus}}, \bibinfo {author}
  {\bibfnamefont{J.~L.}\ \bibnamefont{Bohn}}, \bibinfo {author}
  {\bibfnamefont{J.}~\bibnamefont{Ye}},\ and\ \bibinfo {author}
  {\bibfnamefont{D.}~\bibnamefont{Jin}},\ }%
  \bibfield{journal}{%
  \Doi{10.1103/PhysRevLett.101.080401}{\bibinfo {journal} {Nature Physics}}\ }%
  \textbf{\bibinfo {volume} {7}},\ \bibinfo {pages} {502} (\bibinfo {year}
  {2011})%
  \bibAnnoteFile{NoStop}{miranda2011}%
\bibitem{pasquiou2010}%
  \BibitemOpen
  \bibfield{author}{%
  \bibinfo {author} {\bibfnamefont{B.}~\bibnamefont{Pasquiou}}, \bibinfo
  {author} {\bibfnamefont{G.}~\bibnamefont{Bismut}}, \bibinfo {author}
  {\bibfnamefont{Q.}~\bibnamefont{Beaufils}}, \bibinfo {author}
  {\bibfnamefont{A.}~\bibnamefont{Crubellier}}, \bibinfo {author}
  {\bibfnamefont{E.}~\bibnamefont{Mar\'echal}}, \bibinfo {author}
  {\bibfnamefont{P.}~\bibnamefont{Pedri}}, \bibinfo {author}
  {\bibfnamefont{L.}~\bibnamefont{Vernac}}, \bibinfo {author}
  {\bibfnamefont{O.}~\bibnamefont{Gorceix}},\ and\ \bibinfo {author}
  {\bibfnamefont{B.}~\bibnamefont{Laburthe-Tolra}},\ }%
  \bibfield{journal}{%
  \Doi{10.1103/PhysRevA.81.042716}{\bibinfo {journal} {Phys. Rev. A}}\ }%
  \textbf{\bibinfo {volume} {81}},\ \bibinfo {pages} {042716} (\bibinfo {month}
  {Apr}\ \bibinfo {year} {2010}),\
  \url{http://link.aps.org/doi/10.1103/PhysRevA.81.042716}%
  \bibAnnoteFile{NoStop}{pasquiou2010}%
\bibitem{burt1997}%
  \BibitemOpen
  \bibfield{author}{%
  \bibinfo {author} {\bibfnamefont{E.~A.}\ \bibnamefont{Burt}}, \bibinfo
  {author} {\bibfnamefont{R.~W.}\ \bibnamefont{Ghrist}}, \bibinfo {author}
  {\bibfnamefont{C.~J.}\ \bibnamefont{Myatt}}, \bibinfo {author}
  {\bibfnamefont{M.~J.}\ \bibnamefont{Holland}}, \bibinfo {author}
  {\bibfnamefont{E.~A.}\ \bibnamefont{Cornell}},\ and\ \bibinfo {author}
  {\bibfnamefont{C.~E.}\ \bibnamefont{Wieman}},\ }%
  \bibfield{journal}{%
  \Doi{10.1103/PhysRevLett.79.337}{\bibinfo {journal} {Phys. Rev. Lett.}}\ }%
  \textbf{\bibinfo {volume} {79}},\ \bibinfo {pages} {337} (\bibinfo {month}
  {Jul}\ \bibinfo {year} {1997}),\
  \url{http://link.aps.org/doi/10.1103/PhysRevLett.79.337}%
  \bibAnnoteFile{NoStop}{burt1997}%
\bibitem{Note1}%
  \BibitemOpen
  \bibinfo {note} {Similar behavior has been observed for the $g^2$ function.}%
  \bibAnnoteFile{Stop}{Note1}%
\bibitem{Note2}%
  \BibitemOpen
  \bibinfo {note} {The stability region may grow with increasing temperature
  [14,15]
  %\ref{bisset2011, bisset2012}.
}%
  \bibAnnoteFile{Stop}{Note2}%
\end{thebibliography}
\end{document}